\documentclass[a4paper,%
   a4paper,%
   BCOR12mm,%
   12pt,%
   abstracton,%
   pointednumbers,%
   tablecaptionabove,%
   footinclude,%
   halfparskip,%
   ]{scrartcl}
\usepackage[ilines]{scrpage2} 
\usepackage{epsfig}
\usepackage{multicol}
\usepackage{amsmath}
\usepackage[round,colon,sort,authoryear]{natbib}
\newcommand{\UPLB}{University of the Philippines Los Ba\~{n}os}

\title{\Large Automatic Identification of Animal Breeds and Species\\Using Bioacoustics and Artificial Neural Networks}
\author{\large Jaderick P. Pabico, Anne Muriel V. Gonzales,\\
   \large Mariann Jocel S. Villanueva and Arlene A. Mendoza\\
   \normalsize{Research Collaboratory for Advanced Intelligent Systems}\\
   \normalsize{Institute of Computer Science, \UPLB}\\
   \normalsize{College 4031, Laguna}}
\date{}

\lehead{}
\rohead{\small\thepage}
\lofoot{}
\cofoot{}
\rofoot{}

\begin{document}
\maketitle
%\doublespacing

\begin{abstract}
In this research endeavor, it was hypothesized that the sound produced by animals during their vocalizations can be used as identifiers of the animal breed or species even if they sound the same to unaided human ear. To test this hypothesis, three artificial neural networks (ANNs) were developed using bioacoustics properties as inputs for the respective automatic identification of 13 bird species, eight dog breeds, and 11 frog species. Recorded vocalizations of these animals were collected and processed using several known signal processing techniques to convert the respective sounds into computable bioacoustics values. The converted values of the vocalizations, together with the breed or species identifications, were used to train the ANNs following a ten-fold cross validation technique. Tests show that the respective ANNs can correctly identify 71.43\% of the birds, 94.44\% of the dogs, and 90.91\% of the frogs. This result show that bioacoustics and ANN can be used to automatically determine animal breeds and species, which together could be a promising automated tool for animal identification, biodiversity determination, animal conservation, and other animal welfare efforts.
\end{abstract}

%\keywords{artificial neural network, automated breed and species identification, birds, bioacoustics, dogs, frogs}

%\baselineskip=2.0em  % DOUBLE SPACE
%=============================================================================
% Introduction
\section{Introduction}
Identification of animal breeds or species\footnote{For brevity, the term ``species'' is used hencefort throughout the text to mean either ``breeds'' or ``species'' depending on the context without loss of specificity.} is an important method in animal conservation, biodiversity determination, animal welfare efforts, animal breeding, and other human programs that are geared towards production, improvement, protection and conservation. The usual method to identifying animal species is by visual inspection of the animal's anatomical features vis-\'a-vis a published set of standards~\citep{Ford93, Stettenhelm00, Coile05}. Examples of anatomical features inspected are body dimensions, types and colors of coats, and skin covering. This method is usually done by experts in the field, particularly the breeders and the systematists. Identifying the species of birds that migrate to a certain place, for example, requires a tedious bird-watching procedure that is often conducted over several days. To identify the visiting bird species, the observers need to actually see and capture with a high-resolution camera the images of the animals. Obtaining an unobstracted line of sight between the observer and the observed requires proper positioning of the observer at some safe-enough distance, with some observers going to the extent of wearing camouflage so as not to distract the observed. This procedure becomes much more complicated to conduct if the animals to be observed are nocturnals, are perched atop high canopy trees, or are underwater.

The problems with the current method result from its inherrent requirement that the observer must have an unobstracted line of sight with the observed, and with ample enough lighting. This is because light travels along a straight line and can only be detected by human eyes at a certain range of intensities~\citep{Marriott59}. Because humans are predominantly visual in nature~\citep{Thorpe96}, most of its activities rely much on the sense of seeing, using only the other senses for verifying what was seen~\citep{Ernst02}. Identifying an animal, for example, starts from obtaining an image of its anatomical features, either by the naked eye or as captured by a camera system. The physical process of obtaining an image is by facing the sensor (e.g., eye, camera, or other image capturing devices) towards the direction of the light rays that incidentally reflected on the surface of the animal. The identification is optionally verified when the observed animal produces an identifiable vocalization and is heard by the observer. However, when the animal was not seen (i.e., reflected light did not reach the sensor) or was partially occluded (i.e., reflected light was blocked by another object), no image can be obtained and thus, no identification can be performed even if the vocalization was heard by the observer.

In recent years, researchers have directed their efforts towards utilizing animal vocalizations to detect and identify animals~\citep{Clemins02, Clemins05, Agranat09, Shapiro09, Aide13}. This gives rise to the importance of bioacoustics, the study of animal vocalization. Detecting and identifying animals through their vocalizations do not have the problematic inherrent requirements that identifying through visual means have. Sound waves are propagated by their sources to all directions in space  and do not require an unobstructed line of sight as the waves can bounce back from object to object. Additionally, the observer do not have to find where the animals being observed are located to orient its sensor because the ear, or any other listening devices, can sense the sound waves from any direction and position. Because of this, most of biodiversity recording efforts started from what was heard first in the field, rather than what was first seen.

Animals vocalize to communicate information to other animals, whether within the same species or with another species~\citep{Witzany14}. The reasons for communicating, among many others, include (1)~to impress and attract the opposite sex for reproduction purposes, (2)~to declare territorial boundaries, (3)~to identify family members, (4)~to warn others of the presence of a predator, and (5)~to inform others on the location of food source. For example, birds sing and frogs croak to attract potential mates, while dogs bark to warn their human master of the presence of strangers. 

Without the aid of hearing devices, the vocalizations of most animals within the same species appear the same to humans. This is because researchers found that human ears are sensitive to low-frequency sound while most animals are sensitive to high-frequency ones~\citep{Masterton69}. Other researchers have found out that through evolution, the superhearing, mimicking, and jamming capabilities of animals evolved as a result of survival pressure from their predators~\citep{Barber07, Conner12, Barber13, Igic15}. However, because humans are the ultimate predator, human hearing has evolved to best detect sounds created by other human beings only~\citep{Nelken05}. As a result, the respective sounds produced by two animals from within the same genus but of different species do not appear distinctive from each other with regards to human ear. This is the reason why for the longest time, vocalization was not considered by researchers as a primary identifier of species. However, because of the recent advances in sound technology coupled with the development of complex signal processing methods, it has been found that communication for the purpose of individual identification within the same animal species is possible~\citep{Slabberkoorn02, Janik06, Moore06, Shapiro09, Witzany14}.

The use of vocalization as identifier of species is based on the framework that acoustic properties of sound are used by animals to encode their identities. The spectral features of sound waves, including the fundemantal ones such as frequency and harmonics, differ between two different sources. Because of this, \citet{Cortopassi00} were able to differentiate pairs of orange-fronted parakeets ({\em Aratinga canicularis}) using the  spectral features of bird calls. Similarly, the mother-pup pairs of South American fur seals ({\em Arctocephalus australis}), Galapagos fur seals ({\em Arctocephalus galapagoensis}), and Galapagos sea lions ({\em Zalophus californianus} wollebaeki) were also differentiated via their calls' spectral features~\citep{Trillmich81, Phillips00}. It is based on this framework that the following null ($\mathcal{H}_0$) and alternate ($\mathcal{H}_1$) hypotheses are stated for this effort:
\begin{itemize}
\item[$\mathcal{H}_0$:] There does not exist any combination of the spectral properties of the vocalizations of animals that can differentiate breeds or species with at least 70\% accuracy.
\item[$\mathcal{H}_1$:] There does exist a combination of the spectral properties of the vocalizations of animals that can differentiate breeds or species with at least 70\% accuracy.
\end{itemize}

The minimum accuracy level on the above stated hypotheses was chosen as 20\% more than a randomized toss of coin, i.e., an unbiased coin toss is 50\% while the additional 20\% is attributed to the confidence that the system will provide. The choice of 70\% is arbitrary because the standard for an acceptable accuracy level has not been set by the discipline nor by the industry.

In the past, the process of differentiating objects based on a complex combination of their visual features was automated by a machine vision system (MVS) that uses an artificial neural network (ANN) for efficient classification. ANNs are abstract models of the human brain that is capable of supervised learning and then forming a generalization from a set of experiences~\citep{McCulloch43, Rosenblatt58}. ANNs, for example, have been used by local researchers to automate the error-prone object classification capabilities of humans such as grading the ripeness of tomatoes ({\em Lycopersicon esculentum})~\citep{deGrano07a, deGrano07b}, and differentiating cracks, bloodstains and dirt in eggs~\citep{Zarsuela07}. All of these automated systems used simple cameras as artificial eyes to extend the seeing capabilities of the human eyes beyond their normal working time and way above their normal working rate. As a result, the object classification process was made faster, almost error-free, and used less resources. This improves the classification efficiency significantly over the manual ones~\citep{Pabico08, Pabico09, Pabico12}.

In this effort, the spectral properties of vocalization of three animals were automatically identified using ANNs. The properties were extracted following the various techniques developed in the bioacoustic and the signal processing disciplines. Some of these properties are Spectral Centroid (SC), Spectral Flux (SF), Spectral Roll-off Frequency (SRF), Zero Crossing Rate (ZCR), Mel-Frequency Cepstral Coefficients (MFCC), and Linear Predictive Coding (LPC)~\citep{McEnnis05}. Combinations of these properties were used to train three independent ANNs to identify 13~bird species, eight dog breeds, and 11~frog species, respectively. These animals were chosen not only because they are the most common animals in the Philippines, but the Philippine mountains and forests play hosts to most known migratory species of birds, and are home to vast and various groups of amphibians and reptiles. In fact, the country ranks fourth in the world in terms of bird endemism and first in terms of amphibians and reptiles~\citep{Ong02}. Dogs, on the other hand, were included because Filipinos in general are dog lovers~\citep{Gosling10}. To avoid the Type III errors in classification, each ANN was trained using a 10-fold cross validation technique~\citep{Mosteller48, Kohavi95}.

An ANN that uses a combination of all known 28~spectral properties as its input yielded the highest accuracy rate of 71.43\% for identifying bird species. On the other hand, the ANN that was trained to identify frog species also uses the 28~spectral properties and yielded a 90.91\% accuracy rate. The ANN for identifying dog breeds only required a combination of 4~spectral properties to yield a high accuracy rate of 94.44\%. These results show that there does exist a combination of the spectral properties of the vocalizations of animals that can differentiate breeds or species with high accuracy rate. One implication of this result is that a smartphone technology-enabled ``crowdsourcing'' solution to automatically and transparently collect information on fauna biodiversity by the common people may become a possibility in the near future. This could significantly enhance data collection, not only for biodiversity inventory efforts, but for other animal conservation and welfare endeavors.

%=============================================================================
% Methodology
\section{Materials and Methods}
\subsection{Collecting audio clips of animal vocalizations}
\subsubsection{Vocalization from birds}
Vocalization data from 13~bird species was used in this study. The basic information on these bird species are shown in Table~\ref{tab:1}. Each bird species had 25~different audio clips,  all came from the database of Michigan State University's Avian Vocalizations Center~\citep{AVoCet08}. The total number of audio clips used was~325.

\begin{table}
\caption{The bird species used in vocalization identification showing the common name, scientific name, and place of origin. }\label{tab:1}
\begin{tabular}{rlll}
\hline\hline
 & Common Name & Scientific Name & Place of Origin\\
\hline
 1. & Bananquit                 & {\em Coereba flaveola}        & South America\\
 2. & Black Crake               & {\em Amaurornis ﬂavirostra}   & Africa\\
 3. & Black Hornbill            & {\em Anthracoceros malayanus} & Southeast Asia\\
 4. & Eurasian Skylark          & {\em Alauda arvensis}         & Europe, Asia, Africa\\
 5. & European Goldfinch        & {\em Carduelis carduelis}     & Europe\\
 6. & Philippine Bulbul         & {\em Hypsipetes philippinus}  & Mindanao\\
 7. & Philippine Bush Warbler   & {\em Cettia seebohmi}         & Luzon\\
 8. & Philippine Drongo Cuckoo  & {\em Surniculus velutinus}    & Mindanao\\
 9. & Water Pipit               & {\em Anthus spinoletta}       & Europe, Asia\\
10. & Rufous-tailed Hummingbird & {\em Amazilia tzacatl}        & America\\
11. & Rusty Breasted Cuckoo     & {\em Cacomantis sepulcralis}  & Southeast Asia\\
12. & Spotted Kingﬁsher         & {\em Actenoides lindsayi}     & Luzon\\
13. & White Rumped Shama        & {\em Copsychus malabaricus}   & Southeast Asia\\
\hline\hline
\end{tabular}
\end{table}

\subsubsection{Vocalization from frogs}
Sound data from 11~frog species were collected from different Internet databases~\citep{AmphibiaWeb00}. The species are {\em Boophis luteus}, {\em Bufo marinus}, {\em Fejervarya limnocharis}, {\em Hylarana glandulosa}, {\em Kaloula baleata}, {\em Kaloula pulchra}, {\em Microhyla butleri}, {\em Odorrana hossi}, {\em Phrynoidis aspera}, {\em Polypedates leucomystax} and {\em Rana catesbeiana}. There were a total of 110 audio samples equally distributed to the 11~species.

\subsubsection{Vocalization from dogs}
Bark recordings of beagle, chihuahua, chowchow, shih tzu, and poodle were manually obtained from reputable pet shops. Bark recordings of labrador retriever and syberian husky came from various contributed audio repositories from previous experiments~\citep{Riede99, Jones05}. Ten samples of dog barks audio segments were used for each breed for a total of 90~samples.

\subsubsection{Recordings of negative examples}
Negative examples are sound clips that were neither produced by birds, dogs, nor frogs. These clips are important part of the dataset so that the ANN will be able to differentiate, not only between species, but sound from other sources as well~\citep{Dietterich83, Aha91}. In this research effort, pseudo ``species'' or ``breed'' was added to each dataset. That is, a ``pseudo species'' was added for the bird dataset, a ``pseudo breed'' for the dog dataset, and a ``pseudo species'' for the frog dataset. Thus, the bird, dog and frog datasets have 14~species, 9~breeds, and 12~species, respectively.

\subsubsection{Training, test, and evaluation datasets}
The dataset ($\Sigma$) collected was divided into three sets namely, training ($\Sigma_{\rm train}$), test ($\Sigma_{\rm test}$), and evaluation ($\Sigma{\rm eval}$) sets. If there are a total of $N$ samples in $\Sigma$, then $N$ is as defined in Equation~\ref{eqn:1}, where $N_{\rm train}$ is the total number of samples in the training set, $N_{\rm test}$ is the total number of samples in the test set, and $N_{\rm eval}$ is the total number of samples in the evaluation set. Note that $\Sigma$ is as defined in Equation~\ref{eqn:2} and that the sets are pairwise disjoint such that the expressions in Equations~\ref{eqn:3}, \ref{eqn:4} and~\ref{eqn:5} hold. The fraction of the datasets are $N_{\rm train} = 0.7N$, $N_{\rm test} = 0.1N$, and $N_{\rm eval} = 0.2N$, with all the species equally distributed for each fraction.

\begin{eqnarray}
N &=& N_{\rm train} + N_{\rm test} + N_{\rm eval}\label{eqn:1}\\
\Sigma &=& \Sigma_{\rm train} \bigcup \Sigma_{\rm test} \bigcup \Sigma_{\rm eval}\label{eqn:2}\\
\{ \} &=& \Sigma_{\rm train} \bigcap \Sigma_{\rm test}\label{eqn:3}\\
\{ \} &=& \Sigma_{\rm train} \bigcap \Sigma_{\rm eval}\label{eqn:4}\\
\{ \} &=& \Sigma_{\rm test} \bigcap \Sigma_{\rm eval}\label{eqn:5}\\
\end{eqnarray}

\subsection{Extraction of spectral properties of sound waves}

The spectral properties of the respective audio clips were automatically extracted using a software system called jAudio~\citep{McEnnis05}. These spectral properties, including their respective physical principles, mathematical derivations, and proofs, were discussed in detail elsewhere~\citep{Bogert63,Proakis07} and are not presented here. However, for the benefit of the lay readers as well as for completeness, a brief description of each property is provided as follows:

\begin{enumerate}
\item Mel-Frequency Cepstal Coefficients (MFCC) - Represents the short-term power spectrum of a sound and is primarily used in speech recognition.
\item Zero Crossing (ZC) - The number of times that the time domain signal crosses zero within a given window.
\item Root Mean Square (RMS) - Calculated per window in order to get the amplitude of the sound signal.
\item Fraction of Low Energy Window Frames (FLWEF) - Indicates the variability of the amplitude of windows.
\item Spectral Flux (SF): Signiﬁes the degree of change of the spectrum between windows and is the spectral correlation between adjacent windows.
\item Spectral Rolloff (SR): Indicates the skew of the frequencies present in a window. Eighty five percent of the energy in the spectrum is below this frequency.
\item Compactness (C): Indicates the noisiness of the signal by getting the summation of frequency bins of Fast Fourier Transform (FFT).
\item Method of Moments (MoM): Composed of the ﬁrst ﬁve statistical models that make up the shape of the spectrograph of a given window. The components are area (zeroth order), mean (ﬁrst order), Power Spectrum Density (second order), Spectral Skew (third order), and Spectral Kurtosis (fourth order).
\item Linear Predictive Coding (LPC): Calculates the linear predictive coefﬁcients of a signal in which a particular value is estimated by a linear function of the previous values.
\item Spectral Centroid (SC): Indicates where the ``center of mass'' of the spectrum is and measures the brightness of the sound. This is used as an automatic measure of timbre.
\item Beat Sum (BS): Indicates the sum of the beats in a sound and pertains to the alternating constructive and destructive interference caused by sound waves of different frequency.
\item Strongest Beat (SB): The value of the beat with the strongest frequency.
\item Strength of Strongest Beat (SSB): The intensity of the strongest beat.
\item Spectral Variability (SV): Measures how the ranges of the elements of the sound differ from each other.
\end{enumerate}

The overall average and standard deviation of each of these properties were used as quantified inputs to the ANN. This results to a total of 28 quantified properties as identifiers for species.

\subsection{Structuring, Training and Evaluating ANNs}

\subsubsection{Structuring ANNs}
ANN is an abstract mathematical model of the biological (specifically human) brain composed of a directed network of simple functions and coefficients. Here, the functions are nodes of the network and the coefficients are weights on the edges of the network. A function $f_1$ is connected to another function $f_2$ via a weighted edge represented by the coefficient $\alpha_{1\rightarrow 2}$. Here, the output of $f_1$ is multiplied to coefficient $\alpha_{1\rightarrow 2}$ and their product becomes an input to $f_2$. The direction of the network in this study goes from the quantified spectral properties as inputs to the final species identity as output. The identity is encoded as a computable value represented by an $n$-bit binary system for an $n$-species identification problem.

The general structure of the ANN is that the nodes are generally structured into $m+2$ layers, where the layers are classified into three major classifications: one input layer $L_0$, $m\ge 1$ hidden layers $L_1, L_2, \dots, L_m$, and one output layer $L_{m+1}$. Nodes within a layer are not connected to each other. Here, $L_0$ is composed of at most 28 nodes representing the 28 quantified spectral properties. The output layer is composed of $n$ nodes, where $n - 1$ is the number of species to identify. The extra node is for identifying the negative examples, i.e., the ``pseudo species'' or ``pseudo breed.'' Each of the $m$ hidden layers is composed of the same number of nodes $k$. Nodes in $L_i$ are connected to nodes in $L_{i+1}$, where the sub-graph induced by the nodes in $L_i$ and $L_{i+1}$ form a fully-connected weighted bipartite network with $\alpha_{\{i\}\rightarrow\{i+1\}}$, $\forall i$ as weights. The functions in the nodes are sigmoidal ones $f_{i+1}:\{O_i\} \times \{\alpha_{\{i\}\rightarrow\{i+1\}}\} \rightarrow \{0, 1\}$, where $f_{i+1}$ are the functions in $L_{i+1}$, and $O_i \in \{0, 1\}$ are the outputs of $f_i$ in $L_i$.

The ANNs in this work are denoted as ${\rm ANN}(j, [k, m], n)$, where $j$ is the number of nodes in $L_0$, $k$ is the number of nodes in $L_m$'s, and $n$ is the number of nodes in $L_{m+1}$. The ANN for birds, dogs, and frogs are structured as ${\rm ANN}_{\rm bird}(j, [14, 1], 14)$, ${\rm ANN}_{\rm dog}(j, [9, 1], 9)$, and ${\rm ANN}_{\rm frog}(j, [12, 1], 12)$, respectively. The optimal values of $j$ for each ANN was obtained using a stepwise forward substitution method following the minimum description length criterion for model selection~\citep{Hansen01}.

\subsubsection{Training ANNs}
Each of the ANNs were trained using a feed-forward, back-propagation training algorithm over the samples in $\Sigma_\mathrm {train}$. The errors generated by the initial random assignment for the set of coefficients $\{\alpha\}$ are minimized by back-propagating the identification differences, i.e., ANN classifies an audio clip as having vocalized by species $\mathcal{A}$ but was actually vocalized by species $\mathcal{B}$. The ANN error used for training was the mean square error (MSE). The process was iteratively conducted until one of the following stopping criteria was met: 
\begin{enumerate}
\item When the ANN MSE over the samples in $\Sigma_\mathrm{test}$ has already worsen over several iterations.
\item When the ANN MSE over the samples in $\Sigma_\mathrm{train}$ has not been improved over 100 iterations.
\item When the ANN MSE over the samples in $\Sigma_\mathrm{train}$ has already reached below 0.01.
\end{enumerate}

The whole process described above was repeated nine more times in a 10-fold cross-validation manner~\citep{Mosteller48, Kohavi95}, where at each fold, the samples in each of the sets were changed. For example,  $\sigma_i\not=\sigma_{i+1}$, where $\sigma_i \in \Sigma_{\mathrm{eval},i}$, $\sigma_{i+1} \in \Sigma_{\mathrm{eval},i+1}$, $\forall i$ and $\Sigma_{\mathrm{eval},i}$ is the evaluation set at the $i$th fold.

\subsubsection{Evaluating the ANNs}
The respective accuracies of the ANNs were evaluated over the samples in Σeval. Accuracy is simply computed as the percentage of correctly identified species. An additional secondary evaluation metric called error rate was computed following a multi-class confusion matrix~\citep{Kohavi95, Bavaud06, Manning08}. It is possible for each of the ANNs to have an accuracy of 100\% and a non-zero error rate.

%=============================================================================
% Results and Discussion
\section{Results and Discussion}
\subsection{Collection of audio clips}
The total file sizes of the audio clips collected are 1,100.0~MB, 88.3~MB, and 494.1~MB for birds, frogs, and dogs, respectively. All audio clips were encoded using the Waveform audio file format. This format was used because the spectral properties of soundwave are preserved when the audio stream is stored as raw and uncompressed bitstream.

\subsection{Spectral properties of vocalizations}

\begin{figure}[htb]
\centering\epsfig{file=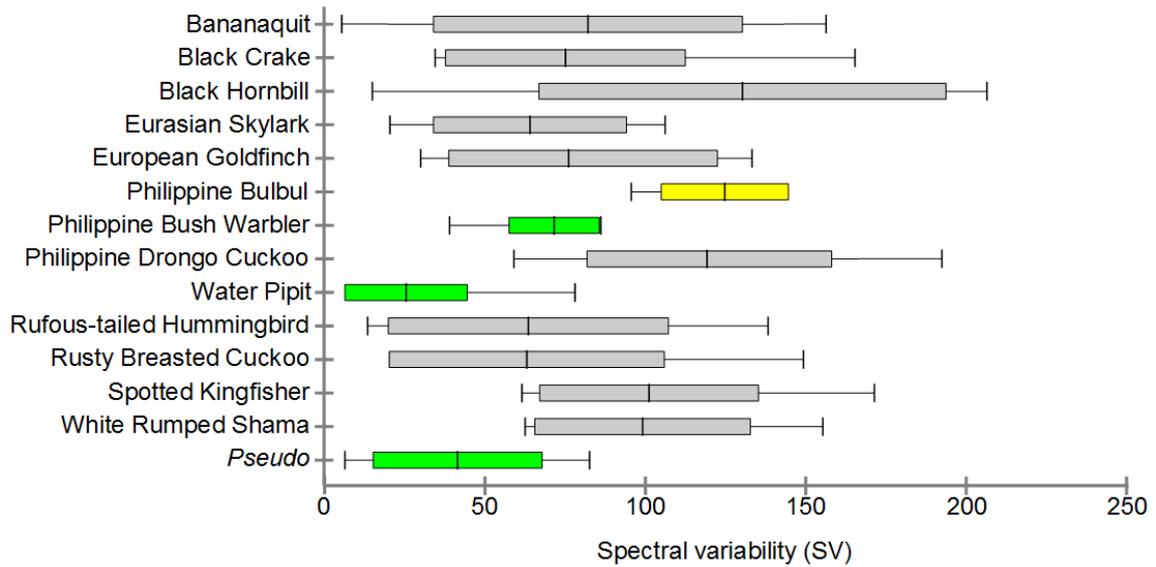, width=6in}
\caption{Box-and-whisker plot of the spectral variability (SV) of bird species. The yellow plot can be visually differentiated from the green plot.}\label{fig:1}
\end{figure}

Figure~\ref{fig:1} shows the SV box-and-whisker plot of each bird species. The plot shows that the respective SV distributions of the Philippine Bulbul and the Philippine Bush Warbler are extremely skewed to the right, while that of the Water Pipit and the Rusty Breasted Cuckoo are extremely skewed to the left. The plot also indicates that SV alone can be a good identifier between Philippine Bulbul and Philippine Bush Warbler, between Philippine Bulbul and Water Pipit, and between Philippine Bulbul and the Pseudo species. The reason for this is that the minimum SV of Philippine Bulbul is greater than the respective maximum SVs of Philippine Bush Warbler, Water Pipit, and Pseudo species. SV alone, however, can not be used to differentiate among Philippine Bush Warbler, Water Pipit, and Pseudo species. In fact, any other pairwise bird species will not be identified by SV alone. Although SV can not differentiate all other bird species, it is just but one dimension in the 28-dimensional spectral properties, the right combination of which may differentiate all pairwise combination of bird species considered in this study. Because of space limitation, the remaining 27 spectral properties for birds, as well as the spectral properties of both dogs and frogs will not be described in this paper.

\subsection{ANN structure}
The respective stepwise forward substitution methods applied to determine the optimal  combination of the 28 spectral properties to structure the $\mathrm{ANN}_\mathrm{bird}$, $\mathrm{ANN}_\mathrm{dog}$, and $\mathrm{ANN}_\mathrm{frog}$ resulted into:

\begin{eqnarray}
\mathrm{ANN}_\mathrm{bird} & = & \mathrm{ANN}(28, [14, 1], 14)\label{eqn:6}\\
\mathrm{ANN}_\mathrm{dog}  & = & \mathrm{ANN}(4, [9, 1], 9)\label{eqn:7}\\
\mathrm{ANN}_\mathrm{frog} & = & \mathrm{ANN}(28, [12, 1], 12)\label{eqn:8}
\end{eqnarray}

This means that it will take a combination of all the 28 quantified spectral properties to differentiate with high accuracy the bird species and frog species, respectively. For dogs, however, only a combination of four spectral properties suffices to differentiate breed barks with high accuracy. These spectral properties are: (1)~Average MFCC, (2)~average MoM, (3)~MoM standard deviation, and (4)~average LPC.

\subsection{Accuracy of ANNs}

Tables~\ref{tab:2}, \ref{tab:3}, and~\ref{tab:4} show the confusion matrices for $\mathrm{ANN}_\mathrm{bird}$, $\mathrm {ANN}_\mathrm{dog}$, and $\mathrm{ANN}_\mathrm{frog}$, respectively.

\begin{table}[htb]
\caption{Confusion matrix for the identification of 13 bird species. Blank entries mean zero.}\label{tab:2}
\centering\begin{tabular}{|l|*{14}{c|}c|}
\hline\hline
\multicolumn{1}{|c|}{Bird}   & \multicolumn{14}{c|}{Bird species as determined by $\mathrm{ANN}_\mathrm{bird}$} & \multicolumn{1}{c|}{Total}\\\cline{2-15}
\multicolumn{1}{|c|}{Species}& a &b &c &d &e &f &g &h &i &j &k &l &m &n &\multicolumn{1}{c|}{Correct} \\ 
\hline\hline
a. Bananaquit          & 5 &  &  &  &  &  &  &  &  &  &  &  &  &  &  5\\\hline
b. Black Crake         &   &3 &  &  &  &  &  &  &1 &  &  &  &  &1 &  3\\\hline
c. Black Hornbill      &   &  &2 &  &  &1 &  &  &  &  &  &1 &  &1 &  2\\\hline
d. Eurasian Skylark    &   &  &  &5 &  &  &  &  &  &  &  &  &  &  &  5\\\hline
e. E. Goldfinch        &   &  &  &  &4 &  &  &  &1 &  &  &  &  &  &  4\\\hline
f. Phil. Bulbul        &   &  &  &  &  &4 &1 &  &  &  &  &  &  &  &  4\\\hline
g. Phil. Bush Warbler  &   &  &  &  &1 &1 &1 &  &  &1 &  &  &1 &  &  1\\\hline
h. Phil. D. Cuckoo     & 1 &  &  &  &  &  &  &4 &  &  &  &  &  &  &  4\\\hline
i. Water Pipit         &   &  &  &  &  &  &  &  &5 &  &  &  &  &  &  5\\\hline
j. R.-t. Hummingbird   &   &  &  &  &  &  &1 &  &1 &3 &  &  &  &  &  3\\\hline
k. R. B. Cuckoo        &   &  &  &  &  &  &  &2 &  &  &3 &  &  &  &  3\\\hline
l. S. Kingﬁsher        &   &  &  &  &  &  &  &  &  &  &1 &4 &  &  &  4\\\hline  
m. W. R. Shama         &   &2 &1 &  &  &  &  &  &  &  &  &  &2 &  &  2\\\hline
n. Pseudo              &   &  &  &  &  &  &  &  &  &  &  &  &  &5 &  5\\\hline\hline
Total Error            &1  &2 &1 &0 &1 &2 &2 &2 &3 &1 &1 &1 &1 &2 & 50\\\hline
\multicolumn{15}{|r|}{Overall error rate (\%)} & 28.57\\\hline
\multicolumn{15}{|r|}{Overall accuracy (\%)}   & 71.43\\\hline\hline
\end{tabular}
\end{table}

The $\mathrm{ANN}_\mathrm{bird}$ was able to identify 100\% of the Eurasian Skylark, without additionally identifying other species as Eurasian Skylark (i.e., its error rate for Eurasian Skylark is 0\%). Bananaquit was also identified 100\% but mistaken one of the Philippine Drongo Cuckoo as a Bananaquit (i.e., error rate of 1.82\%). All of Water Pipits were identified 100\% as well, but mistakenly identified one black crake, one European Goldfinch, and one Rufous-tailed Hummingbird as Water Pipits. Thus, the $\mathrm{ANN}_\mathrm{bird}$ has an error rate of 5.45\% for identifying Water Pipit. The Pseudo species was also correctly identified 100\%, but also identifed Black Crake and Black Hornbill as Pseudo species (i.e., error rate of 3.64\%). Only one of the samples was identified by $\mathrm{ANN}_\mathrm{bird}$ as Philippine Bush Warbler (i.e., accuracy of 20\%). It also identified one Philipine Bulbul and one  Rufous-tailed Hummingbird as  Philippine Bush Warbler (i.e., error rate of 3.64\%). The overall accuracy of $\mathrm{ANN}_\mathrm{bird}$ is 71.43\% while its overall error rate is 28.57\%. With regards to identifying bird species, the null hypothesis $\mathcal{H}_0$ is rejected and the alternate hypothesis $\mathcal{H}_1$ is accepted in its stead.

The $\mathrm{ANN}_\mathrm{dog}$ was able to identify all dog breeds correctly (i.e., 100\% accuracy, Table~\ref{tab:3}). It correctly identified the Pseudo breed 50\% of the time. The other half of the Pseudo breed was identified erroneously as Chow Chow (i.e., error rate of identifying Chow Chow is 5.56\%). The error rate for identifying all the other breeds, including the Pseudo breed is 0\%. The overall accuracy of $\mathrm{ANN}_\mathrm{dog}$  is 94.44\% and its overall error rate is 5.56\%. With these results, the alternate hypothesis $\mathcal{H}_1$ is accepted instead of the null hypothesis $\mathcal{H}_0$.

\begin{table}[htb]
\caption{Confusion matrix for the identification of eight dog breeds. Blank entries mean zero.}\label{tab:3}
\centering\begin{tabular}{|l|*{9}{c|}c|}
\hline\hline
\multicolumn{1}{|c|}{Dog} & \multicolumn{9}{c|}{Breed according to $\mathrm{ANN}_\mathrm{dog}$} & \multicolumn{1}{c|}{Total}\\\cline{2-10}
\multicolumn{1}{|c|}{Breed} & a &b &c &d &e &f &g &h &i &\multicolumn{1}{c|}{Correct} \\ 
\hline\hline
a. Beagle             & 2 &  &  &  &  &  &  &  &  &  2\\\hline
b. Chihuahua          &   &2 &  &  &  &  &  &  &  &  2\\\hline
c. Chow Chow          &   &  &2 &  &  &  &  &  &  &  2\\\hline
d. Labrador Retriever &   &  &  &2 &  &  &  &  &  &  2\\\hline
e. Pomeranian         &   &  &  &  &2 &  &  &  &  &  2\\\hline
f. Poodle             &   &  &  &  &  &2 &  &  &  &  2\\\hline
g. Shih Tzu           &   &  &  &  &  &  &2 &  &  &  2\\\hline
h. Siberian Husky     &   &  &  &  &  &  &  &2 &  &  2\\\hline
i. Pseudo             &   &  &1 &  &  &  &  &  &1 &  1\\\hline
Total Error           & 0 &0 &1 &0 &0 &0 &0 &0 &0 & 17\\\hline\hline
\multicolumn{10}{|r|}{Overall error rate (\%)} & 5.56\\\hline
\multicolumn{10}{|r|}{Overall accuracy (\%)} & 94.44\\\hline\hline
\end{tabular}
\end{table}

All frog species (Table~\ref{tab:4}), except for the Pseudo species were identified by $\mathrm{ANN}_\mathrm{frog}$ correctly (i.e., 100\% accuracy for each species and 0\% accuracy for the Pseudo species). $\mathrm{ANN}_\mathrm{frog}$ also identified all Pseudo species as {\em B. luteus}. Thus, $\mathrm{ANN}_\mathrm{frog}$ has an error rate of 9.09\% for identifying {\em B. luteus}. $\mathrm{ANN}_\mathrm{frog}$ has an overall accuracy of 90.91\% and an overall error rate of 9.09\%. The accuracy of $\mathrm{ANN}_\mathrm{frog}$ proves that the alternate hypothesis $\mathcal{H}_1$ must be accepted and the null hypothesis $\mathcal{H}_0$ rejected.

\begin{table}[htb]
\caption{Confusion matrix for the identification of 11 frog breeds. Blank entries mean zero.}\label{tab:4}
\centering\begin{tabular}{|l|*{12}{c|}c|}
\hline\hline
\multicolumn{1}{|c|}{Frog} & \multicolumn{12}{c|}{Frog species as determined by $\mathrm{ANN}_\mathrm{frog}$.} &\multicolumn{1}{c|}{Total}\\
\cline{2-13}\multicolumn{1}{|c|}{Species} & a &b &c &d &e &f &g &h &i &j &k &l & \multicolumn{1}{c|}{Correct} \\ 
\hline
a. \em{B. luteus}           & 2 &  &  &  &  &  &  &  &  &  &  &  &  2\\\hline
b. \em{B. marinus}          &   &2 &  &  &  &  &  &  &  &  &  &  &  2\\\hline
c. \em{F. limnocharis}      &   &  &2 &  &  &  &  &  &  &  &  &  &  2\\\hline
d. \em{H. glandulosa}       &   &  &  &2 &  &  &  &  &  &  &  &  &  2\\\hline
e. \em{K. baleata}          &   &  &  &  &2 &  &  &  &  &  &  &  &  2\\\hline
f. \em{K. pulchra}          &   &  &  &  &  &2 &  &  &  &  &  &  &  2\\\hline
g. \em{M. butleri}          &   &  &  &  &  &  &2 &  &  &  &  &  &  2\\\hline
h. \em{O. hossi}            &   &  &  &  &  &  &  &2 &  &  &  &  &  2\\\hline
i. \em{P. aspera}           &   &  &  &  &  &  &  &  &2 &  &  &  &  2\\\hline
j. \em{P. leucomystax}      &   &  &  &  &  &  &  &  &  &2 &  &  &  2\\\hline
k. \em{R. catesbeina}       &   &  &  &  &  &  &  &  &  &  &2 &  &  2\\\hline
l. Pseudo                   & 2 &  &  &  &  &  &  &  &  &  &  &0 &  0\\\hline
Total Error                 &2  &0 &0 &0 &0 &0 &0 &0 &0 &0 &0 &0 & 20\\\hline\hline
\multicolumn{13}{|r|}{Overall error rate (\%)} & 9.09\\\hline
\multicolumn{13}{|r|}{Overall accuracy (\%)} & 90.91\\\hline\hline
\end{tabular}
\end{table}

%=============================================================================
% Summary and Conclusion
\section{Summary and Conclusion}
Three ANNs for respectively identifying 12 bird species, eight dog breeds, and 11 frog species were structurally optimized, trained, and evaluated. The animals were identified through the 28 quantifiable spectral properties of their respective vocalizations. Identifying bird and frog species requires all 28 properties while identifying dog breeds only requires four, namely average MFCC, average and standard deviation of MoM, and  average LPC. The respective accuracies of identifying bird species, dog breeds, and frog species are 71.43\%, 94.44\%, and 90.91\%. The respective alternate hypotheses that there exists combinations of the spectral properties of the vocalizations of these three animals that can differentiate among their respective breeds or species with at least 70\% accuracy are accepted. Thus, by using bioacoustics methods, the sound produced by animals during their vocalizations can be used as identifiers of species, and that the process can be automated by ANNs.

One of the implications of this result is that species inventory procedures for biodiversity recording purposes maybe augmented and enhanced through a technique called ``crowdsourcing''~\citep{Estelles12}. This technique is similar to the Christmas Day Bird Census that was started in 1900 and has become an annual tradition by the members of the National Audubon Society~\citep{Robbins15}. In crowdsourcing, however, the participants are not only bird enthusiasts but common people equipped with a modern-day smartphones whose applications can automatically record animal vocalizations, identify the animals, and send the dated and geo-located information to a central database. This automated process maybe transparently done by the application without user intervention, which is a much better system than what~\citet{Huetz12} proposed.

%=============================================================================
\section*{Acknowledgements}

The following persons and entities who provided information and data are instrumental to this reseach effort: (1)~Dr. Juan Carlos T. Gonzales, Director of UPLB Museum of Natural History who provided information on bird vocalizations; (2)~Dr. Arvin Diesmos, Herpetologist of the National Museum who provided information on frogs; and (3)~the Cartimar Pet Center and the Pet City Hobby Specialist for the dog bark recordings.

%=============================================================================

% Need a baselineskip=15pt so that reference section is not double spaced
%\baselineskip=15pt
%\parskip=20pt
\bibliography{bioacoustics}
\bibliographystyle{plainnat}

%\balancecolumns
\end{document}